\documentclass[letterpaper]{jacow}
\usepackage{pdfpages,ragged2e}
\usepackage{xcolor}
\begin{document}
\title{Obvious and non-obvious aspects of digital Self-Excited-Loops
for SRF cavity control}
\author{Larry Doolittle\thanks{LRDoolittle@lbl.gov},
Shreeharshini Murthy, LBNL; \\
Matei Guran, Lennon Reyes, Shrividhyaa Sankar Raman,
Philip Varghese, FNAL}
\maketitle

\begin{abstract}
In 1978, Delayen showed how Self-Excited Loops (SEL) can be used to great advantage for controlling narrow-band SRF cavities. Its key capability is establishing closed-loop amplitude control early in the setup process, stabilizing Lorentz forces to allow cavity tuning and phase loop setup in a stable environment.
As people around the world implement this basic idea with modern FPGA DSP technology, multiple variations and operational scenarios creep in that have both obvious and non-obvious ramifications for latency, feedback stability, and resiliency.
This paper will review the key properties of a Delayen-style SEL when set up for open-loop, amplitude stabilized, and phase-stabilized modes. Then the original analog circuit will be compared and contrasted with the known variations of digital CORDIC-based implementations.
\end{abstract}

\section{Physics}

The state-space representation for cavity voltage $V$ is
$$ {dV\over dt} = aV + bK + cI$$
where (following Heaviside) all of $V$, $K$, $I$, $a$, $b$, and $c$
are complex numbers, varying slowly compared to the timescale of the
RF resonance itself.
We ignore the units of $V$ and $K$ in this discussion, except to note
that $b$ and $K$ have to be self-consistent.
$a$ necessarily has units of s$^{-1}$.
This discussion will also ignore the beam-loading term $cI$.

SRF cavities have the confusing property that $a$ is not constant;
as their sheet-metal construction bends with the acoustic environment
and Lorentz forces, the imaginary part of $a$ (detuning) varies in real time.
That means that the governing equation is not LTI (Linear Time Invariant),
although for short time scales that's still a useful approximation.

One of Delayen's key 1978\cite{JDthesis} contributions was to start from
the usual Lorentzian equation for a resonator output as a function of detuning
$$ {V\over K} = {1\over 1 + j\chi}~~~,$$
where $\chi = \Im(a) / \Re(a) $.  That can be inverted to give the needed drive
$$ K = V (1 + j\chi)~~~.$$
A single feedback loop for the reactive component of the drive
signal will therefore ``fix'' both amplitude and phase fluctuations created
by detuning.  His diagram for the requisite hardware is shown.

\begin{figure}[!htb]
    \centering
    \includegraphics*[width=0.7\columnwidth]{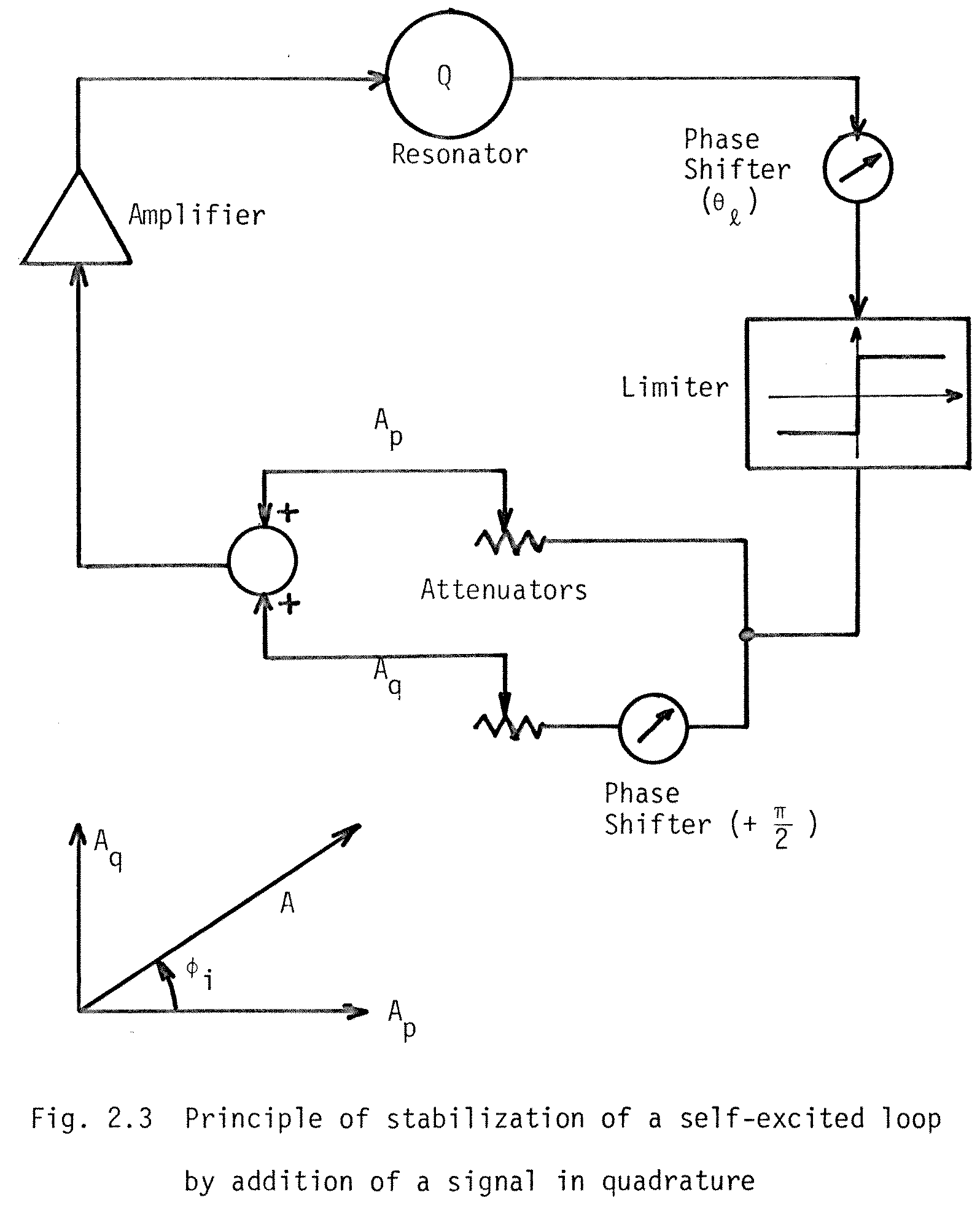}
\end{figure}

The observed benefit in that era of very narrow-band helix resonators
was that the field could be brought to its operating point without
regard for fine frequency tuning.  Even the amplitude feedback loop can be
engaged while detuned; with the amplitude loop closed, potential ponderomotive
instabilities are strongly suppressed\cite{USPAS}.
Finally, once the resonance is adjusted to the operational value,
the tuning loop can be closed.
While amplitude and phase loops should be closed (not clipping) during
beam operation, temporary frequency excursions that clip the phase feedback loop
will naturally return to lock without manual or automated intervention.
More on that later.

\section{Abstract SEL behavior and feedback}

\begin{figure}[!htb]
    \centering
    \includegraphics*[width=0.9\columnwidth]{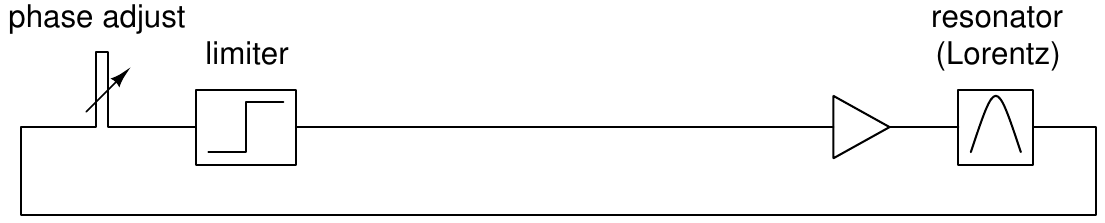}
    \caption{Textbook SEL oscillator}
\end{figure}

\begin{figure}[!htb]
    \centering
    \includegraphics*[width=0.9\columnwidth]{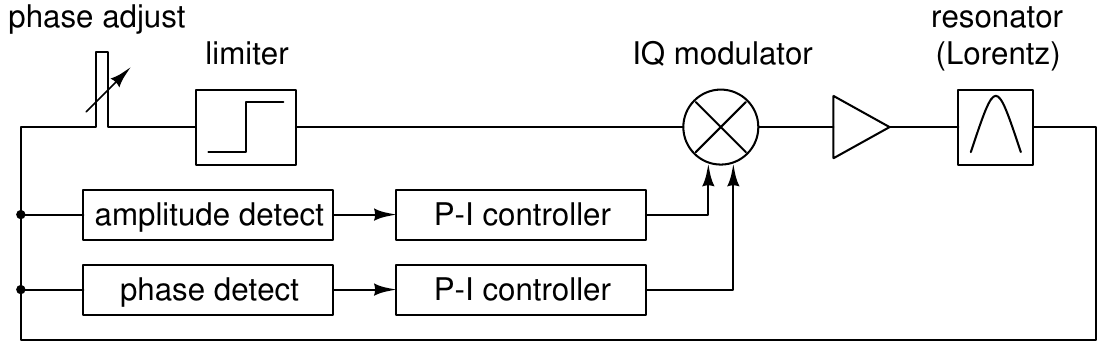}
    \caption{Delayen 1978 cavity controller}
\end{figure}

A traditional analog SEL oscillator has a phase adjustment
to compensate for cable lengths to give pure positive feedback.
The resulting system ``starts from noise.''
After start-up, the limiter limits, and the amplitude feedback gain
drops to \hbox{0}. Phase feedback literally has a gain of +1.00.

Delayen added amplitude and phase stabilization loops to that core design.
This stiff negative PI feedback totally overwhelms that baseline SEL ``positive feedback.''

\section{CORDIC}

The CORDIC structure was invented by Jack Volder in 1959\cite{Volder} to compute
trigonometric functions using pure shift-and-add digital hardware.
Every such CORDIC block produces
$$ z_{\rm out} = A \cdot z_{\rm in}\prod_{i=0}^{n-1} \exp\left(j\sigma_i\theta_i\right)$$
$$ \theta_{\rm out} = \theta_{\rm in} - \sum_{i=0}^{n-1} \sigma_i\theta_i$$
\vskip -2ex
where
$$ \theta_i = \tan^{-1} 2^{-i} $$
$$ A = \prod_{i=0}^{n-1} \sqrt{1+2^{-2i}} \approx 1.64676 $$
There are two standard mechanisms for choosing $\sigma_i$.

Rotation mode: $\theta_{\rm out} \approx 0$, used to get
$z_{\rm out} = A \cdot z_{\rm in} \cdot \exp\left(j\theta_{\rm in}\right)$.
In the special case when $z_{\rm in} = 1/A$,
that yields $\cos\theta_{\rm in}$ and $\sin\theta_{\rm in}$.

Vectoring mode: $\Im\left(z_{\rm out}\right) \approx 0$,
used to get $\theta_{\rm out} = \angle z_{\rm in}$ and\hfil\break
$\Re\left(z_{\rm out}\right) = A \cdot |z_{\rm in}|$.

\begin{figure}[!htb]
    \centering
    \includegraphics*[width=0.8\columnwidth]{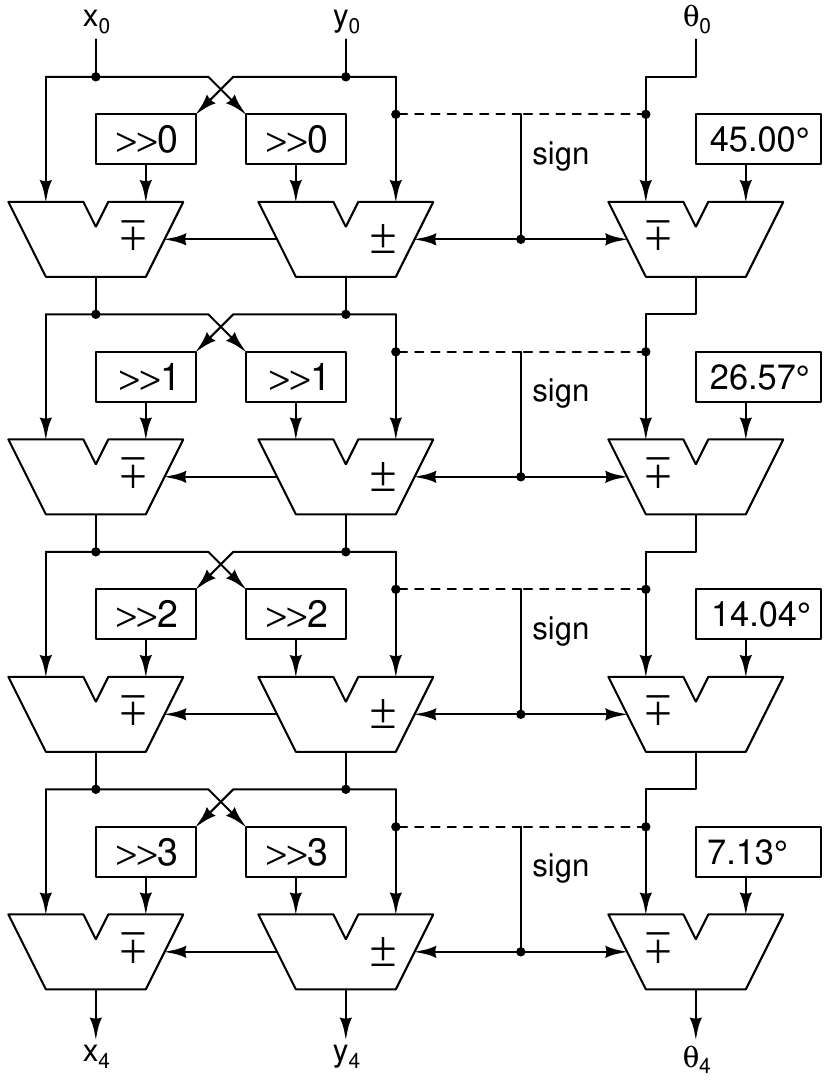}
    \caption{CORDIC internal structure}
\end{figure}

An FPGA implementation of CORDIC will therefore always have three inputs
and three outputs.  We create a visual vocabulary showing that in
rotation mode (note the ``R'' in the CORDIC block), the $\theta$ output
is unused (and is within rounding error of 0).
Similarly, with vectoring mode (marked with ``V'') the $y$ output is ignored.
Additional special cases yield traditional polar-to-rectangular and
rectangular-to-polar configurations.

\begin{figure}[!htb]
    \centering
    \includegraphics*[width=0.7\columnwidth]{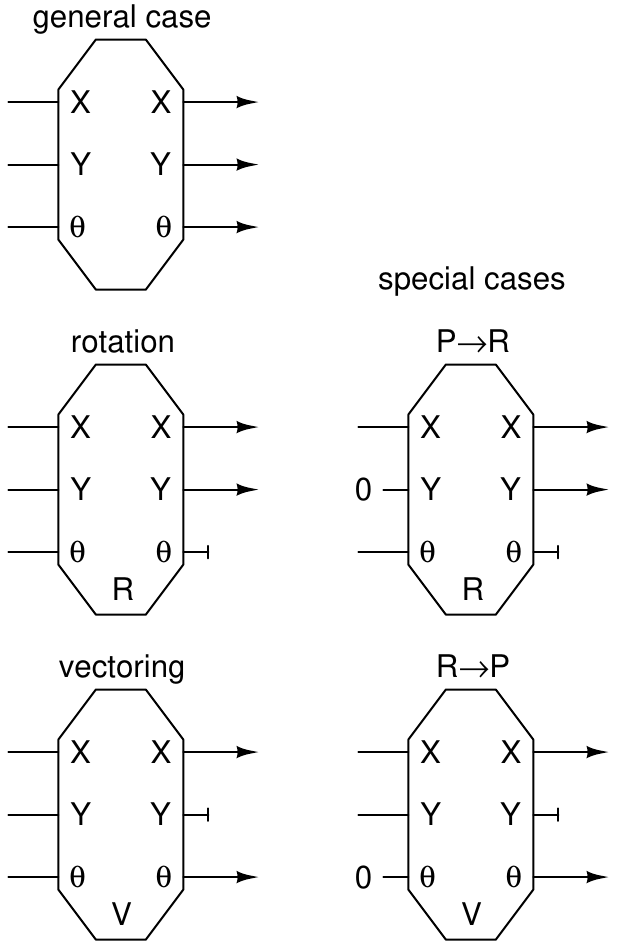}
    \caption{CORDIC use cases}
\end{figure}

The CORDIC module supplied in LBNL's Bedrock code base\cite{Bedrock} is pipelined,
and the selection between vectoring and rotation modes can be made
on a cycle-by-cycle basis, allowing a single CORDIC instance to be
time-multiplexed between different tasks in the final circuit.
See a LLRF 2013 tutorial\cite{Resource}.

\section{PI Controller}

The first rule of PID controllers is that there is no \hbox{D}.
A block diagram of a combined proportional and integral
implementation is shown,
together with the visual vocabulary used here to represent it.

\begin{figure}[!htb]
    \centering
    \includegraphics*[width=0.95\columnwidth]{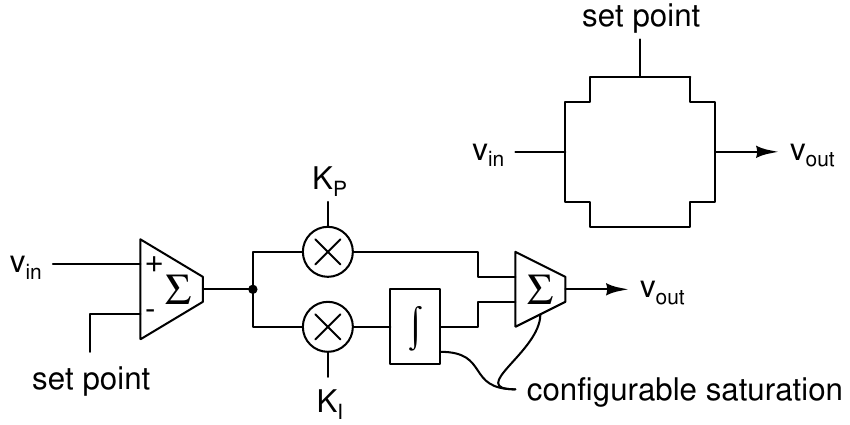}
    \caption{PI controller structure and representation}
\end{figure}

Clipping of SEL phase-tracking loop is essential part of the process!
That is the output of one of these PI controllers; it therefore needs
a zero-windup clipping implementation, as well
as good runtime control of the clip value.

This block diagram shows the I term as gain followed by
the (saturating) integrator; the reverse doesn't make sense when there
is a possibility of run-time gain adjustments.
This circuit can be ``reset'' (force output and integrator state to zero)
as needed by setting the saturation level to zero.

\section{Considerations for off-frequency operation}

Simple phase detectors can create unpleasant results when faced with
off-frequency operation.  Clues for how to properly mitigate that comes
from the AD9901 chip (1996)\cite{AD9901}.  In the SEL context, the output of the
phase setpoint subtracter should be processed with a state machine.
When the actual phase error wraps around from positive to negative,
for instance, that condition needs to be registered, so the output
of the phase detector stays saturated at the maximum positive value.
The LBNL code base refers to this as a Stateful Phase Resolver.

\begin{figure}[!htb]
    \centering
    \includegraphics*[width=0.85\columnwidth]{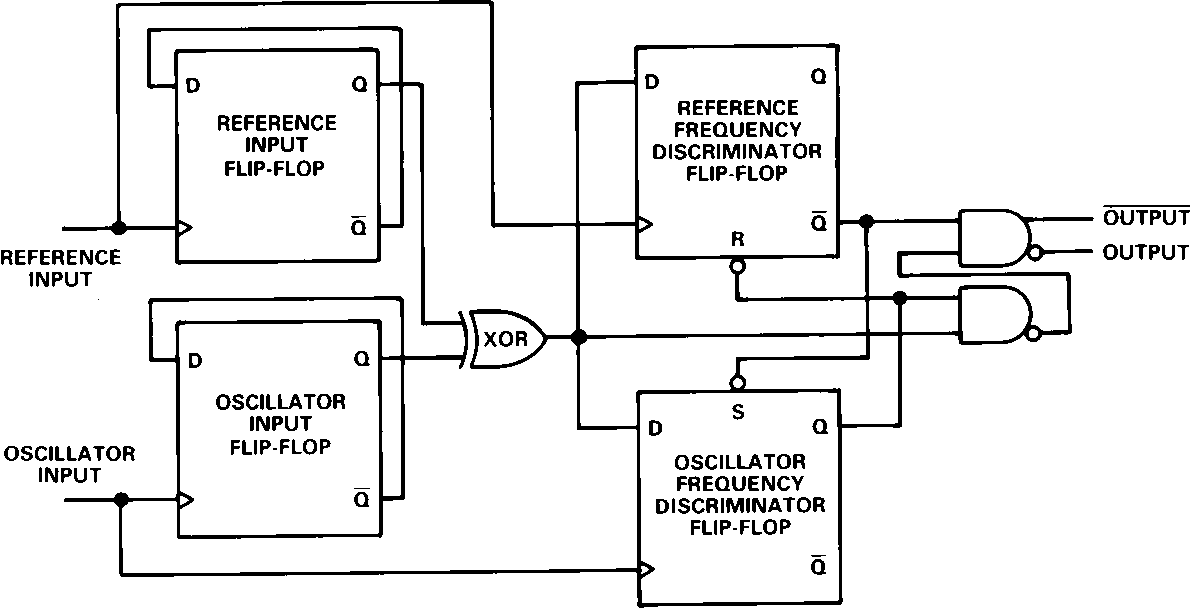}
    \caption{AD9901 Functional Block Diagram}
\end{figure}

When such a phase resolver is used, the reactive drive created by
the phase loop will ``stick'' at that limit until the frequency error
is removed -- presumably by a tuner control loop not discussed here.

\section{Digital SEL topologies}

A series of topologies for realizing SEL in DSP are shown:
LBNL\cite{CMOC}, JLab\cite{JLab}, S-DALINAC\cite{SDALINAC} and \hbox{BARC}\cite{BARC}.  The GDR topology is shown for completeness.
Each diagram assumes $x$ and $y$ (a.k.a. I and Q) inputs and outputs.
Digital and/or analog down- and up-conversion, required to complete
the LLRF system, are off-topic here.

Although details vary, they all do share the core 1978-vintage feature
of detecting phase errors, and using a feedback controller to generate a drive
of form $1+j\chi$ to fix both amplitude and phase errors.

\begin{figure}[!htb]
    \centering
    \includegraphics*[width=0.85\columnwidth]{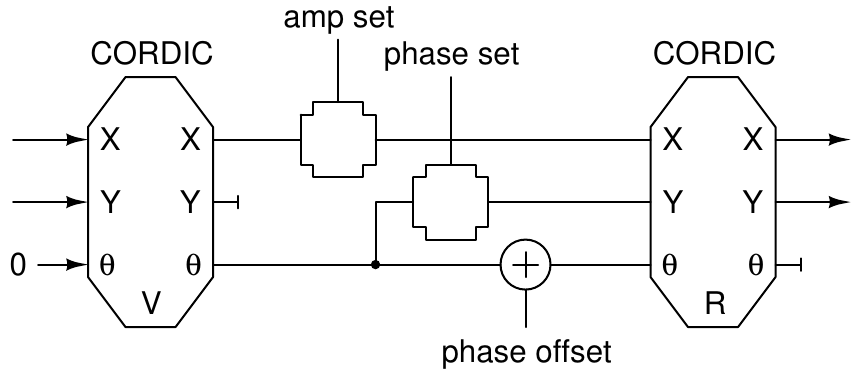}
    \caption{LBNL (2015)}
\end{figure}

The JLab design introduces an odd serialization of
amplitude and phase feedback paths.
The third diagram suggests extending the JLab idea to re-orthogonalize
the amplitude and phase loops.  Now neither PI output is routed through
the second (delay-creating) CORDIC.

The BARC design uses no CORDIC blocks; instead it builds a limiter
out of squaring circuits and a local feedback loop.  Its logic footprint
is heavier on multipliers than the other designs, which is not a concern
for typical FPGAs today.  The lack of CORDIC means its phase feedback latency
is lower than the other designs shown.

\begin{figure}[!htb]
    \centering
    \includegraphics*[width=0.95\columnwidth]{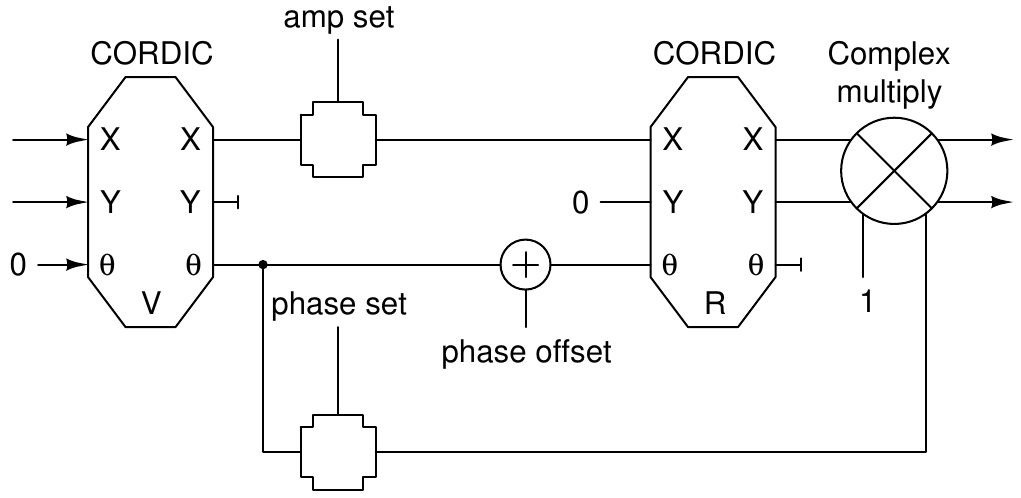}
    \caption{JLab (2008)}
\end{figure}

\begin{figure}[!htb]
    \centering
    \includegraphics*[width=0.95\columnwidth]{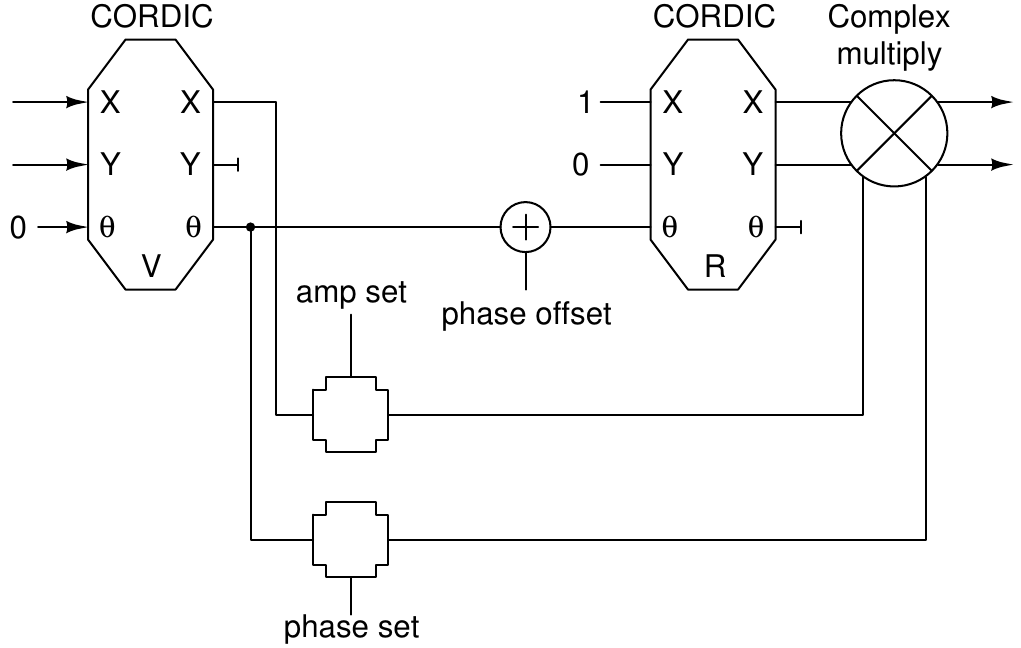}
    \caption{Suggested}
\end{figure}

\begin{figure}[!htb]
    \centering
    \includegraphics*[width=0.95\columnwidth]{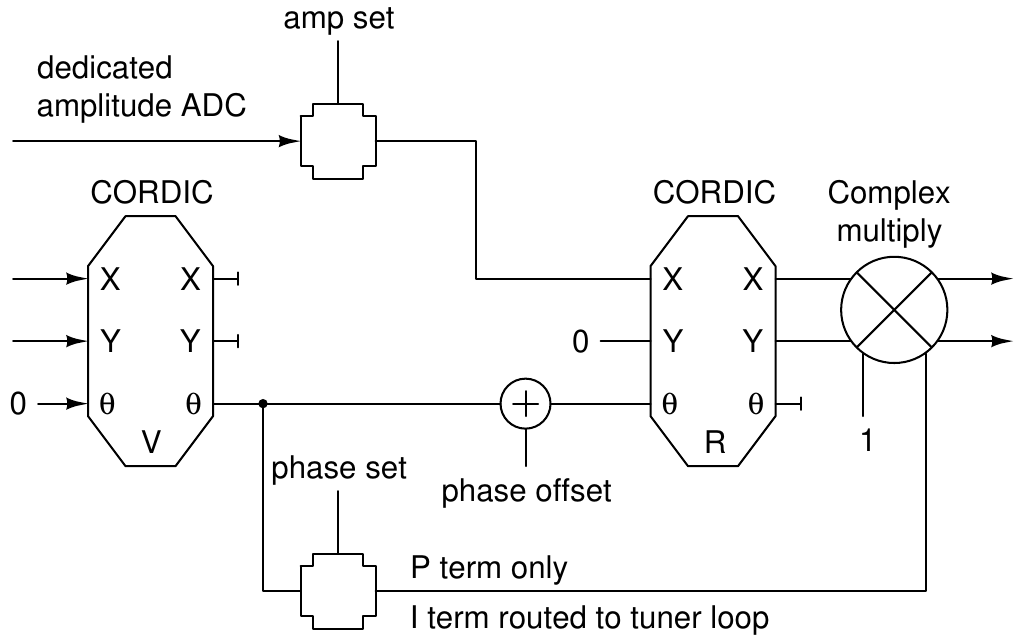}
    \caption{S-DALINAC (2011)}
\end{figure}

\begin{figure}[!htb]
    \centering
    \includegraphics*[width=0.95\columnwidth]{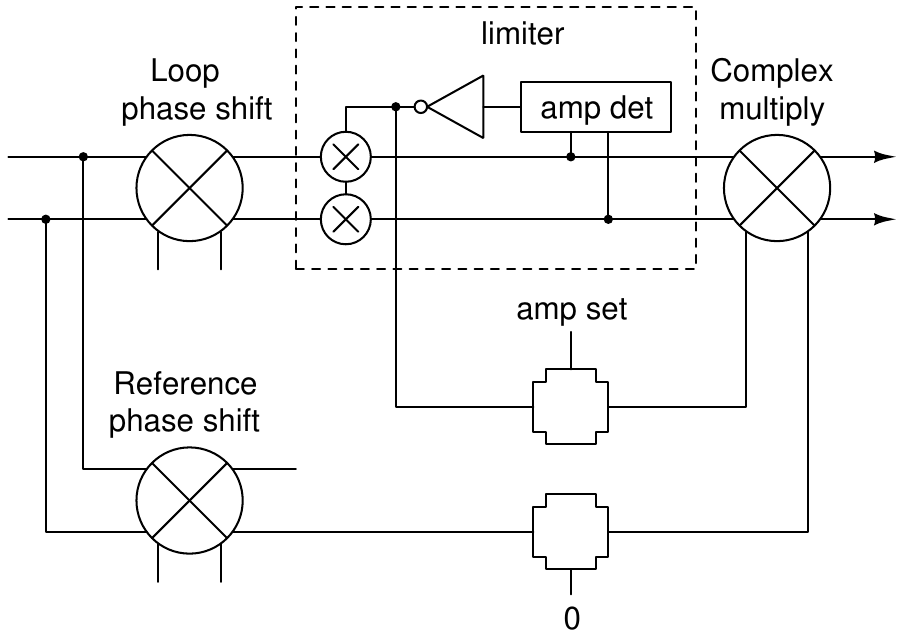}
    \caption{BARC (2014)}
\end{figure}

\begin{figure}[!htb]
    \centering
    \includegraphics*[width=0.95\columnwidth]{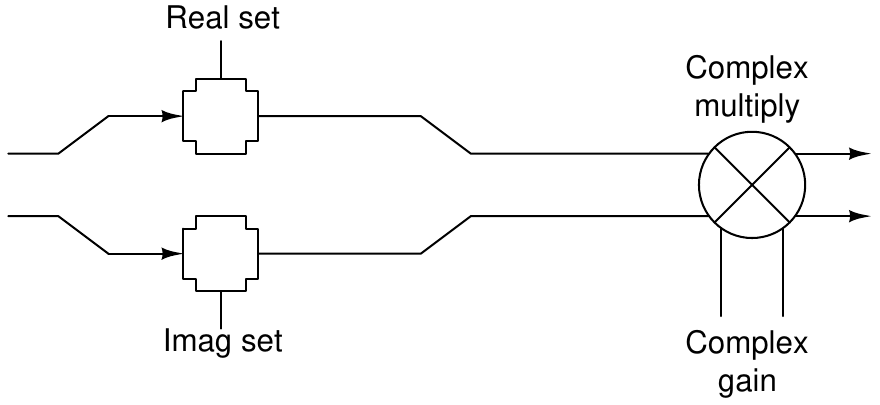}
    \caption{GDR}
\end{figure}

\section{Stability}

As the pole position $a$ moves around in the imaginary axis,
the dynamics of the system also changes.  It's important to understand
if and how that changes the stability of the feedback system.
We assert that if the condition $\Im(a) << \omega_{\rm 0dB}$ is
maintained, the system stability is not materially affected.

\begin{figure}[!htb]
    \centering
    \includegraphics*[width=0.95\columnwidth]{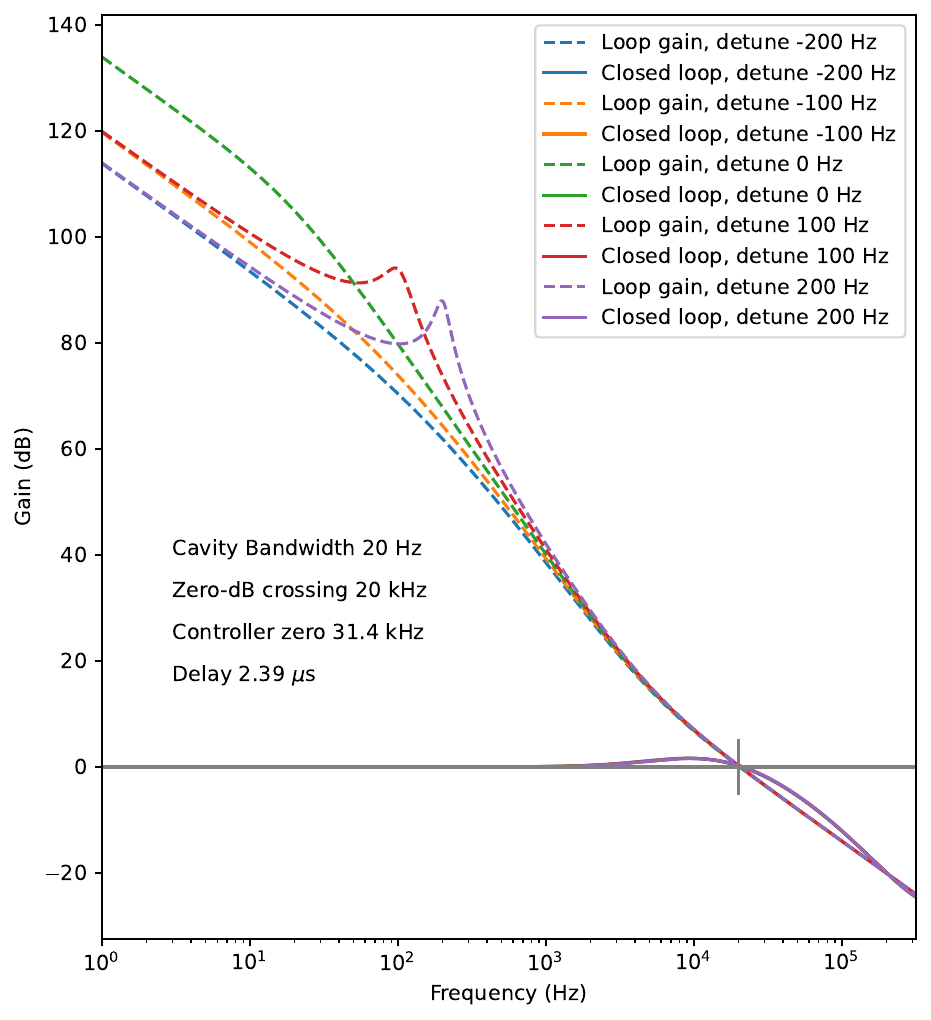}
    \caption{Open- and closed-loop gains with detuning}
\end{figure}

A Nyquist stability graph with example numbers will illustrate that.

\section{Operation}

It's important to understand how an SEL is intended to work
when the resonator frequency deviates from nominal (e.g., microphonics)
and the reactive drive required to stay locked exceeds the defined limits.

The experimental results shown were presented at \hbox{LLRF17}\cite{LCLS2prototype}.
The simulated waveforms use an idealized feedback controller.

\begin{figure}[!htb]
    \centering
    \includegraphics*[width=1.0\columnwidth]{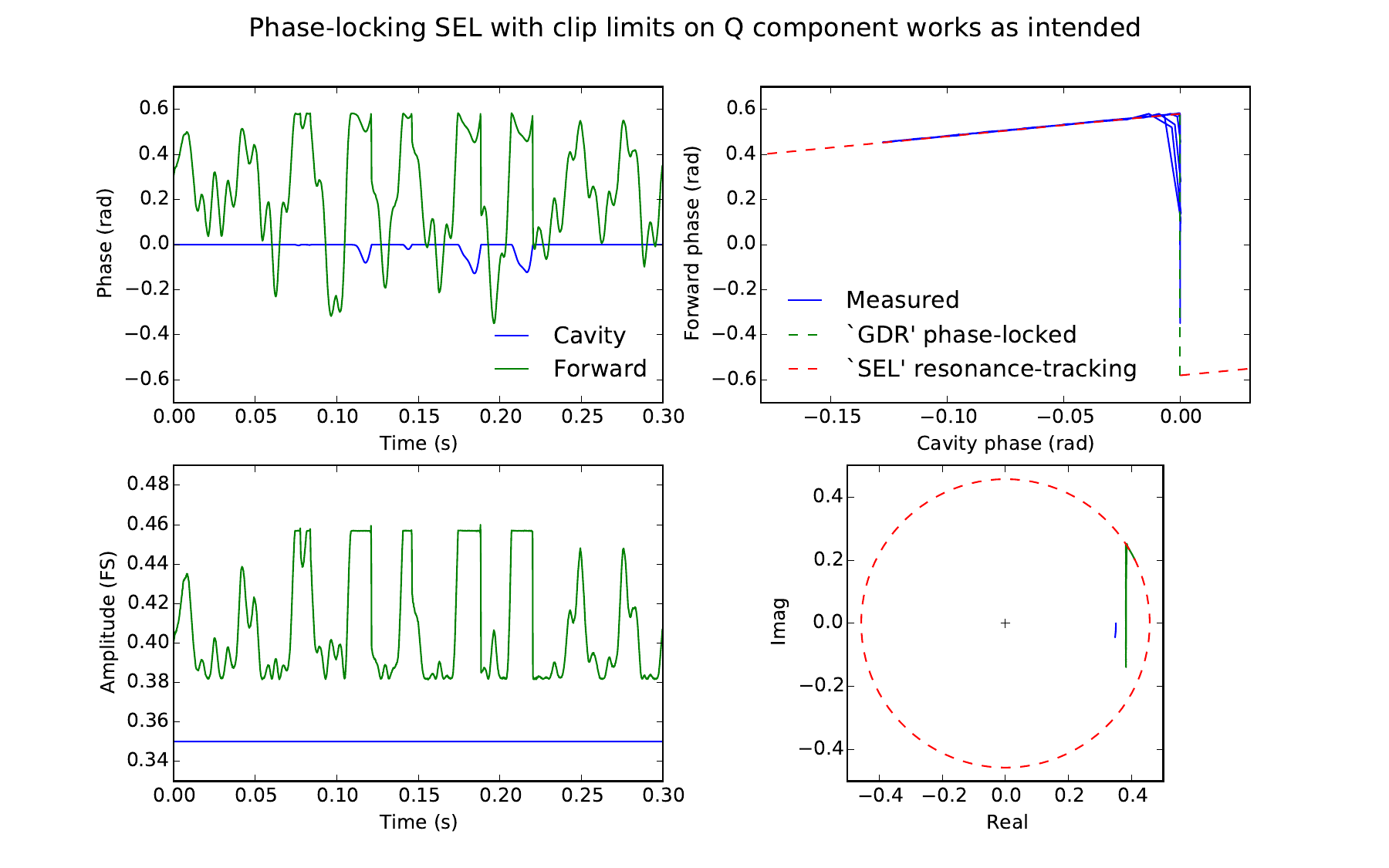}
    \caption{Measured waveforms}
\end{figure}

\begin{figure}[!htb]
    \centering
    \includegraphics*[width=1.0\columnwidth]{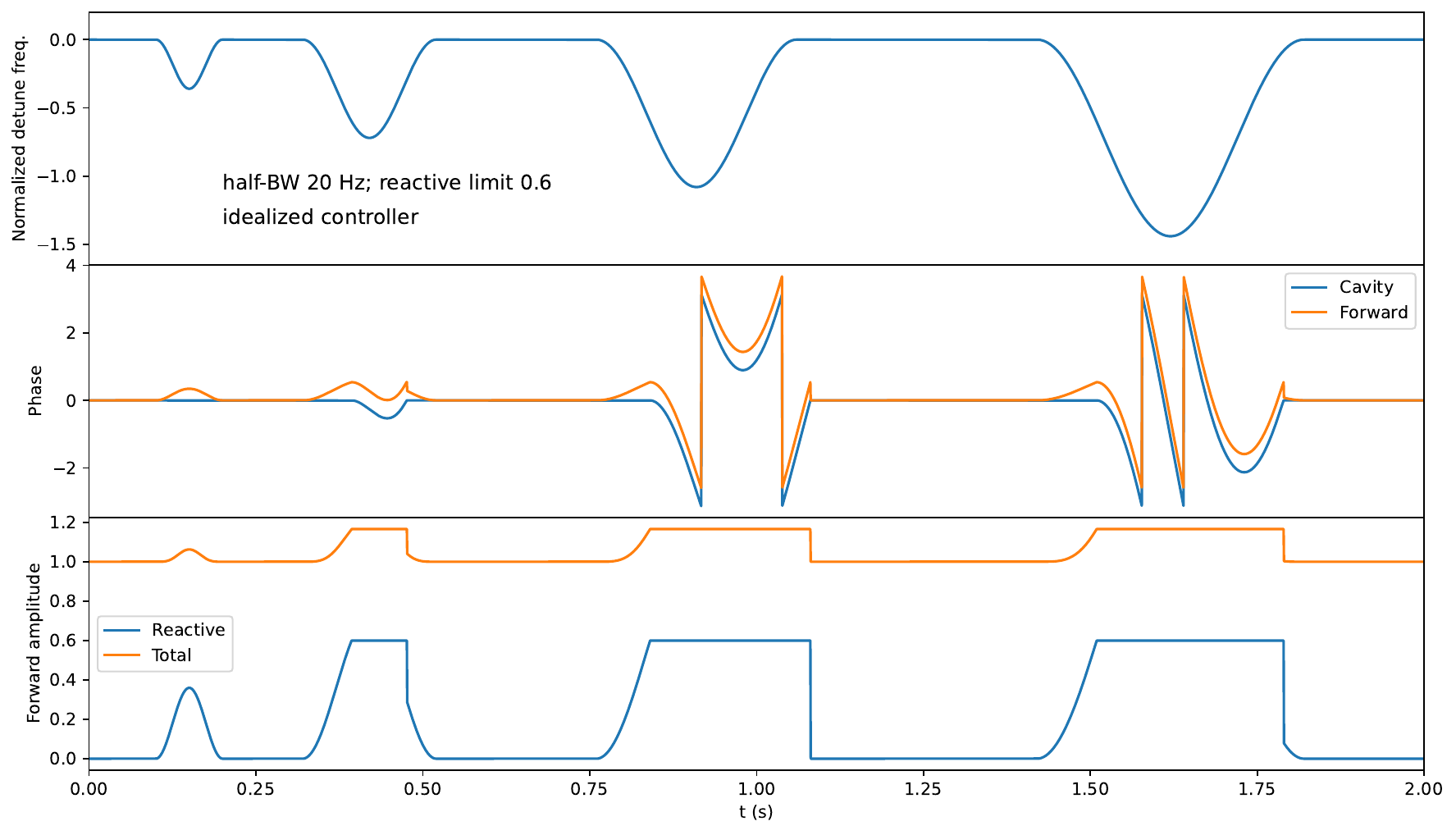}
    \caption{Simulated waveforms}
\end{figure}

\section{Future work}

Future work could attempt to achieve lower latency.
One approach is to simply use higher clock speeds for the DSP;
FPGAs are faster now than they were in the 2008-2014 time frame.
There is also a possibility to add a direct P term, bypassing the
relatively slow CORDIC steps.  That would require careful management
of both its gain coefficients and a fallback configuration
if/when clipping is detected.

Leader/follower CORDIC tricks can easily reduce (by a factor of two)
the latency of the two-CORDICs-in-series phase-following function
in the final suggested SEL topology, without affecting the feedback path.

We wish for more code-sharing, in particular the ability
to mix-and-match test-benches and implementations from labs around the world.
Of course that will require surmounting barriers from licensing
and disparate languages.

\section{Not considered}

\begin{itemize}
\item Units for and calibration of cavity state value
\item System-ID and drift correction, including PI gain setup\cite{Drift}
\item Analog and/or digital down- and up-conversion
\item Mitigation of instabilities caused by high P-gain and nearby cavity passband modes
\item Tx channel linearization.
\item Beam loading corrections (timing-based feedforward)
\end{itemize}



\begin{thebibliography}{99}
\def\etal{{\it et al.}}

\bibitem{JDthesis}
\textit{Phase and Amplitude Stabilization of Superconducting Resonators},
J. Delayen, Ph.D. thesis, 1978, Caltech

\bibitem{USPAS}
\textit{Ponderomotive Instabilities, Microphonics, and RF Control},
J. Delayen, USPAS June 2008

\bibitem{Volder}
\textit{The CORDIC Computing Technique}, J. Volder,
1959 Proceedings of the Western Joint Computer Conference

\bibitem{AD9901}
AD9901 data sheet, Analog Devices, 1996

\bibitem{CMOC}
\textit{Accelerator-On-Chip Simulation Engine},
L. Doolittle \etal, LLRF 2015, Shanghai

\bibitem{JLab}
\textit{Development of a Digital Self-Excited Loop for Field
Control in High-Q Superconducting Cavities},
J. Delayen \etal, SRF2007, Beijing

\bibitem{SDALINAC}
\textit{Design and Commissioning of a Multi-frequency
Digital Low Level RF Control System},
M. Konrad \etal, IPAC 2011, San Sebasti\'an

\bibitem{BARC}
\textit{Digital self-excited loop for a superconducting linac},
G. Joshi \etal, NIM A 2014

\bibitem{Bedrock}
\textit{Bedrock}, https://github.com/BerkeleyLab/Bedrock

\bibitem{Resource}
\textit{Avoiding Resource Overutilization in FPGAs},
L. Doolittle, LLRF 2013, Lake Tahoe

\bibitem{LCLS2prototype}
\textit{LCLS-II LLRF prototype testing and characterization},
L. Doolittle \etal, LLRF 2017, Barcelona

\bibitem{Drift}
\textit{Drift observations and mitigation in LCLS-II RF},
A. Benwell \etal, LLRF 2023, Gyeongju

\end{thebibliography}
\end{document}